\documentclass[seceq]{ptptex}
\usepackage{wrapft}

\usepackage{graphicx}

\newcommand{\bra}[1]{\left\langle #1 \right|}
\newcommand{\ket}[1]{\left| #1 \right\rangle}

\newcommand{\vecr}{\mbox{\boldmath $r$}}

\newcommand{\vectau}{\mbox{\boldmath $\tau$}}

\newcommand{\rhot}{\tilde{\rho}}

\title{Di-neutron correlation in soft octupole excitations of
neutron-rich Ni isotopes beyond $N=50$}

\author{Yasuyoshi \textsc{Serizawa}$^1$ 
\footnote{E-mail: serizawa@nt.sc.niigata-u.ac.jp}
and Masayuki \textsc{Matsuo}$^2$
\footnote{E-mail: matsuo@nt.sc.niigata-u.ac.jp}
}

\inst{
$^1$Graduate School of Science and Technology, Niigata University, Niigata 950-2181, Japan\\
$^2$Department of Physics, Faculty of Science, Niigata University, Niigata 950-2181, Japan
}

\abst{
We investigate low-lying octupole response of neutron-rich
Ni isotopes beyond the N=50 shell closure using the 
Skyrme-Hartree-Fock-Bogoliubov mean-fields and the continuum
quasi-particle random phase approximation. Performing detailed numerical
analyses employing the Skyrme parameter set SLy4 and a
density-dependent delta interaction of the mixed type, we show that
a neutron mode emerges
above the neutron separation energy as a consequence of the weak binding of
neutrons and it exhibits strong influences of the di-neutron correlation. 
}

\begin{document}

\maketitle

\section{Introduction}

The di-neutron correlation, 
a spatial pair correlation with small correlation length
among neutrons,
has been one of the central themes of the physics
of two-neutron halo nuclei such as $^{11}$Li
\cite{Hansen,Esbensen,Ikeda,Sackett,Shimoura,Zinser,Ieki}. 
Although
an affirmative experimental signature
of the neutron spatial correlation in $^{11}$Li is obtained rather recently, \cite{Nakamura}  
 many theoretical
predictions have been accumulated, concerning not only
two-neutron halo 
nuclei\cite{Hansen,Esbensen,Ikeda,Zhukov,Barranco01,Aoyama,Hagino05,Hagino07,Hagino08}
 but also other neutron-rich systems
including the surface area of medium- and heavy- mass neutron rich
nuclei\cite{MMS05,Pillet}
and
dilute neutron matter\cite{Matsuo06,Matsuo07,Margueron08}. Indeed the analysis 
of the spatial structure of the neutron Cooper pair in 
dilute matter\cite{Matsuo06,Matsuo07,Margueron08} has revealed
a mechanism that the di-neutron correlation may emerge generically.
Namely the spatial di-neutron
correlation originates from a strong coupling feature
of the neutron pair correlation, which becomes significant
at low densities because of the strong
momentum dependence of the attractive nuclear force in
$^1S$ channel. Furthermore 
the induced pairing interaction 
caused by the exchange of surface phonons 
is claimed to give additional contribution to possible enhancement
of the spatial correlation in finite nuclei\cite{Pastore}.

The influence of the di-neutron correlation on nuclear structure
could be multifold. Namely it may influence not only  
the ground state but also 
various modes of excitations and dynamics. Attentions, however,  have been
focused so far mostly on soft dipole excitation in two-neutron halo nuclei
\cite{Hansen,Esbensen,Ikeda,Sackett,Shimoura,Zinser,Ieki,Nakamura,Hagino05},
and
rather recently, in medium- and heavy-mass neutron rich nuclei
\cite{MMS05,Matsuo07}.
Concerning the latter case, we have shown in our previous study
\cite{MMS05} that strong influence of the di-neutron correlation  in
the soft dipole excitation of proton semi-magic neutron-rich
Ca and Ni nuclei.
Considering the possible generality of the di-neutron correlation,
it is expected that this correlation may also emerge
in other multipole modes of excitation. 
We anticipate it
in multipole excitations with negative and natural parity 
because the di-neutron correlation 
involves a configuration mixing among single-particle orbits with 
different parities. In the present paper, therefore, we examine low-lying octupole
modes of excitation.  

There exists only a few 
investigations of low-lying octupole excitations 
in neutron-rich nuclei while in stable nuclei the low-lying
$3^-$ state with a character of octupole surface vibration 
is well established\cite{BM2,Spear89}. 
Experimental properties of the low-lying
octupole modes in neutron-rich nuclei are known very
little, with exceptions such as $^{20}$O\cite{KBG00}. On the theoretical side,
it is predicted \cite{Fa91,Sag95,Cat96} that neutron halo nuclei exhibit a neutron mode
with large octupole strength 
in the continuum region above the neutron threshold, resulting from 
transitions from weakly bound orbits
to continuum orbits, and thus having a completely different
character with the surface vibration. 
Analyses based on the random phase approximation (RPA)\cite{HSZ01,ZZD03,Yo01,PWM04},
which can describe both the continuum
states and the collectivity,
 have shown 
coexistence of the collective surface vibration
and neutron continuum strength near the threshold
in medium mass neutron rich nuclei.
Note however that models of Refs.\cite{HSZ01,ZZD03,Yo01,PWM04} 
do not include the pair correlations, and 
analyses are limited to nuclei with doubly closed shell configurations,
such as $^{60}$Ca, $^{28}$O and  $^{68,78}$Ni. 
The quasi-particle random phase
approximation including the pair correlation effect is employed in 
Ref.\cite{KBG00} to
describe oxygen isotopes $^{18-24}$O. There, however, the Hartree-Fock+BCS 
approximation is adopted instead of the Hartree-Fock-Bogoliubov
scheme, and effects of the pair correlation are not
discussed intensively.
In contrast to these
preceding works, we would like to focus in the present paper on
 roles of the pair correlation and the possibility
of the di-neutron correlation especially in the low-lying octupole
correlation. For this purpose
we use the continuum quasi-particle random phase approximation (the
continuum QRPA) \cite{MMS05,Matsuo07,Matsuo01}
which can
describe both the continuum, the pairing, and the collectivities
in neutron rich medium-mass nuclei. 
We choose nickel isotopes $^{80-86}$Ni for numerical analysis,
and perform detailed investigation of $^{84}$Ni which is chosen
as an representative example. A preliminary report of this work
is seen in Ref.\cite{previous}.

\section{Skyrme-HFB plus continuum QRPA method}

\begin{figure}[htbp]
\begin{center}
\resizebox{0.8\linewidth}{!}{\includegraphics{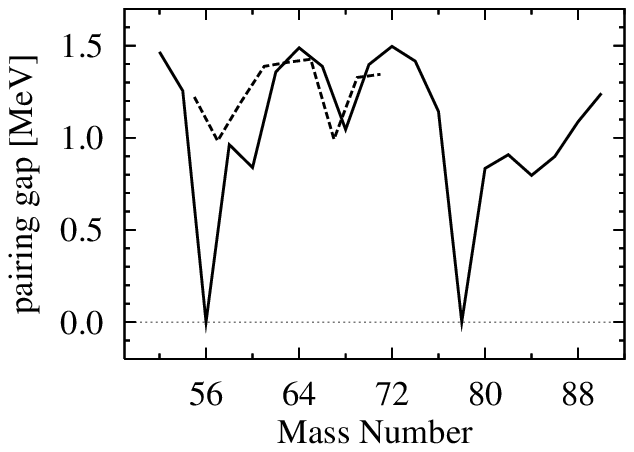}}
\end{center}
\caption{
The average neutron pairing gap of nickel isotopes.
The solid line shows the values obtained in the present HFB calculation using SLy4 
and the mixed-type DDDI.
The dashed line is the experimental value extracted from the nuclear masses
using the three-point formula.
}
\label{pairing}
\end{figure}

In the present study we assume the spherical symmetry of the
ground state as we analyze semi-magic nickel isotopes.
We first construct the spherically symmetric ground state and associated 
self-consistent mean-fields by
means of the coordinate-space Skyrme-Hartree-Fock-Bogoliubov 
(Skyrme-HFB) method\cite{DFT84,DNW96},
with which we can describe properly the spatially extended
wave functions of weakly bound quasi-particle orbits. 
We employ the Skyrme-force parameter set SLy4 \cite{CBH98},
which has been extensively used for studies of the
neutron-rich nuclei \cite{DNR01,DNS02,DN02,SDN03,MDL00,BDP00,TOU03,BHR03}.
We shall also use another parameter set SkM$^*$\cite{BHR03,SKMstar} 
for comparison.
In our previous study of the dipole excitation\cite{MMS05}
a Woods-Saxon potential was adopted, but here we
perform fully self-consistent HFB calculations.
Concerning the effective pair interaction, we 
use the density-dependent
delta-type interaction (DDDI) \cite{Esbensen,DNR01,BHR03} 
which is given by
\begin{equation}\label{DDDI}
v_{pair}(\vecr,\vecr')={1\over2}V_0(1-P_\sigma)
\left(1-\eta{\rho(\vecr)\over \rho_0}\right)\delta(\vecr-\vecr').
\end{equation}
We choose the so-called mixed type pairing,\cite{DN02}
i.e. $\eta=1/2, \rho_0$ = 0.16 fm$^{-3}$, concerning the
density dependence,  and we 
determine the strength of the pair interaction $V_0$ in the same way adopted 
in Ref.\cite{MMS05}. Namely we determined
$V_0 = -315 $MeV fm$^3$ for nickel isotopes
so that the average pairing gap for neutrons is
in overall agreement with the experimental 
odd-even mass difference evaluated
with the three-point formula \cite{SDN98} (cf. Fig.\ref{pairing}).

We  describe excitation modes by means of the continuum QRPA. The
continuum QRPA formalism is essentially the same as those in
Refs.\cite{Matsuo01,MMS05} except that we here formulate it on the
basis of the Skyrme-Hartree-Fock-Bogoliubov mean-fields and different
residual interactions.
Concerning the 
residual interaction in the pairing channel, we adopt the one
derived from the functional derivatives of the pairing energy functional,
as was in our previous study\cite{MMS05}.
Concerning the residual interaction in the particle-hole channel, 
we use the same Skyrme interaction employed in constructing the ground state, but we take
a Landau-Migdal approximation to the Skyrme interaction\cite{KSG02,KSG05,PVKC07}
in order to make the systematic numerical calculations feasible.
In practice, the particle-hole interaction is given by
$v_{ph}(\vecr-\vecr')= 
\left\{(F_0/N_0)(\vecr) + (F'_0/N_0)(\vecr)\vectau\cdot\vectau') \right\}
\delta(\vecr-\vecr')$
where $F_0$, $F_0'$ and $N_0$ are the Landau-Migdal parameters
\cite{Migdal,NO87,BJS75} and the
associated normalization factor, evaluated for the Skyrme interaction.
Their expressions are given, for example, in Refs.\cite{GS81,BDE02}.
The Landau-Migdal parameters are defined usually for symmetric nuclear 
matter, 
but we treat the Fermi momentum $k_F$ in the expressions as a local
quantity $k_F(\vecr)=\left(3\pi^2\rho(\vecr)/2\right)^{1/3}$ related to
the nucleon density $\rho(\vecr)$ \cite{BDE02}. Thus the force strength is density dependent,
and hence
depends on 
the spatial coordinate $\vecr$. 

Given the HFB mean-fields and the residual interactions, we solve
the linear response equation
\begin{equation} \label{rpa}
\left(
\begin{array}{c}
\delta\rho_{q L}(r,\omega) \\
\delta\rhot_{+,q L}(r,\omega) \\
\delta\rhot_{-,q L}(r,\omega) 
\end{array}
\right)
=\int_0 dr'
\left(
\begin{array}{ccc}
& & \\
& R_{0,q L}^{\alpha\beta}(r,r',\omega)& \\
& & 
\end{array}
\right)
\left(
\begin{array}{l}
\sum_{q'}\kappa_{ph}^{qq'}(r')
\delta\rho_{q' L}(r',\omega)/r'^2 + v^{ext}_{q L}(r') \\
\kappa_{pair}(r')\delta\rhot_{+,q L}(r',\omega)/r'^2 \\
-\kappa_{pair}(r')\delta\rhot_{-,q L}(r',\omega)/r'^2
\end{array}
\right).
\end{equation}
to describe the correlated multipole 
response of a nucleus against an external field.
Here $\delta\rho_{q L}(r,\omega)$, $\delta\rhot_{+,q L}(r,\omega)$,
and $\delta\rhot_{-,q L}(r,\omega)$ are  responses in the
normal and abnormal densities with multi-polarity $L$, 
and   $\kappa_{ph}$ and $\kappa_{pair}$ are related to the interaction strengths
$(F_0/N_0)(r), (F'_0/N_0)(r)$ and $V_0(1-\eta\rho(r)/\rho_0)$.
Note that we construct the response function 
$ R_{0,q L}^{\alpha\beta}(r,r',\omega)$
in terms of products of two single-quasi-particle HFB Green's
function summed over not only the discrete quasi-particle states but also
the  continuum quasi-particle states \cite{Matsuo01,MMS05}.
This is possible because we use the exact HFB Green's function\cite{Bel87}
consisting of the regular and
out-going wave solutions of the HFB equation. 
We consider the external field 
\begin{equation}
V^{ext}(\vecr)=\sum_i r_i^3Y_{30}(\hat{\vecr}_i)\ \ \  {\rm and } \ \ \
\sum_i{1 \pm \tau_{3,i} \over 2}r_i^3Y_{30}(\hat{\vecr}_i), \ \ \
\end{equation}
for the isoscalar,  neutron and proton strength functions 
$dB(\lambda L)/dE=\sum_i B(\lambda L;0_{gs}\rightarrow 3_i^-)\delta(E-E_i)$
($\lambda L =$IS3, n3 and p3, respectively),
which can be evaluated as 
$dB(\lambda L)/dE= -{7 \over \pi}{\rm Im} \int \sum_q dr v_{{\rm
ext},q}(r)\delta\rho_{qL}(r,\omega)$ using the solution $\delta\rho_{qL}(r,\omega)$ 
of Eq.(\ref{rpa}).

Numerical details are as follows.
The HFB equation is solved using the radial coordinate  in a spherical
box $r=0-r_{max}$. 
The Skyrme-HFB code is our original one, whose numerical procedures essentially
follow those of Ref.\cite{DFT84}. We solve 
the radial HFB equation using the
Runge-Kutta method instead of the Numerov method adopted in 
Refs.\cite{DFT84,BD05}
since we need a consistent evaluation of 
 derivatives of the quasi-particle wave functions  in
constructing the Green's function used in the continuum QRPA calculation. 
We have checked that results of our Skyrme-HFB code agree with those produced by the
code HFBRAD \cite{BD05}.
The continuum QRPA part is based on our previous version\cite{MMS05} employing the Woods-Saxon
potential, but here we replace the Woods-Saxon potential by the
Skyrme-HFB mean-fields. Also we implement the 
Landau-Migdal approximation of the Skyrme interaction 
as the residual interaction in the particle-hole channel.
As the box size we choose  $r_{max}$ = 22 fm,
and an equidistant discretization with $\Delta r=0.2$ fm.
Concerning the cut-off of the quasi-particle orbits, we set 
$l_{max} = 17 \hbar $ 
for the single-particle partial waves $lj$, and $E_{max}=60$ MeV
with respect to the quasi-particle energies. 
The cut-off energy is
a standard choice adopted in many Skyrme HFB calculations \cite{DNR01,DNS02,SDN03,TOU03,BHR03},
but slightly larger than
that adopted in our previous Woods-Saxon
calculation\cite{MMS05}.
Concerning
the orbital angular momentum cut-off, we find that the ground state and
the octupole strength function have good convergence already at $l= 12
\hbar$, but to evaluate the transition densities we need larger
$l$'s.
Because we adopt the Landau-Migdal approximation, the consistency 
between the ground state and the excited states is partly broken.
In order to minimize effects of the self-consistency breaking, 
we renormalize the strength of the particle-hole residual interaction 
as $v_{ph} \rightarrow f\times v_{ph}$ by a numerical factor $f$ so that 
the spurious center-of-mass mode in the isoscalar dipole excitation
has zero excitation energy.

\section{Numerical analysis}

\subsection{Octupole strength functions in $^{84}$Ni}\label{sec:strength}

We shall first discuss a representative example
$^{84}$Ni 
in order to clarify
basic features of low-lying octupole excitation.

\begin{figure}[htbp]
\begin{center}
\resizebox{0.8\linewidth}{!}{\includegraphics{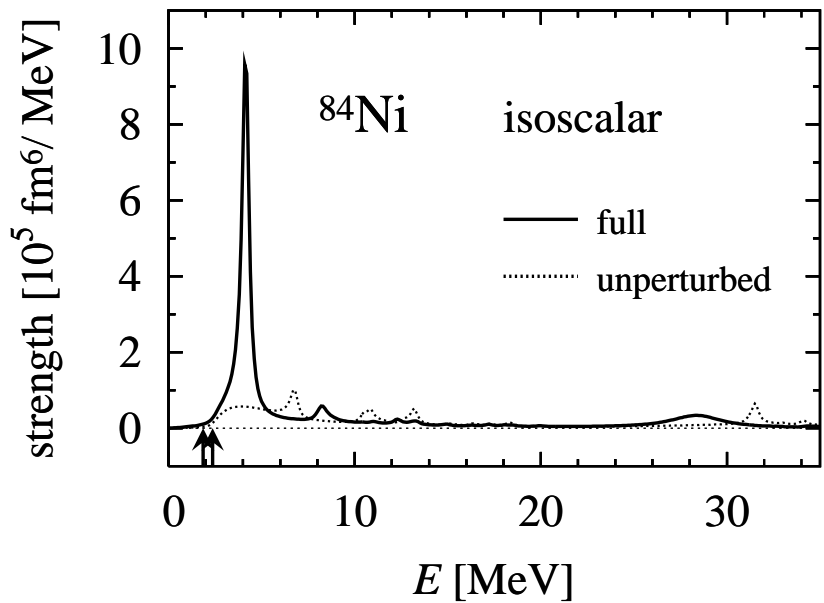}}
\end{center}
\caption{The isoscalar octupole strength $dB({\rm IS3})/dE$,  plotted  with the solid
 curve, obtained for $^{84}$Ni using SLy4 and the mixed-type DDDI.  
The smoothing parameter is $\epsilon=0.2$ MeV. 
The unperturbed strength is also shown with the dotted curve.
The arrows indicate the one- and two-neutron separation energies.
}
\label{strength.is.ni84}
\end{figure}

In Fig.~\ref{strength.is.ni84}, we show the isoscalar octupole strength
in $^{84}$Ni. Here the strength function is calculated with
smoothing parameter $\epsilon = 0.2$ MeV, which is introduced
in the linear response equation as the imaginary part of
the excitation frequency $\omega+ i\epsilon$. 
This means that the strength function is folded with 
an Lorentzian function with FWHM $=2\epsilon=0.4$ MeV.
It is seen that there are essentially two groups of strength distribution;
one around $E$ = 26 - 32 MeV and the low-lying distributions below
$E\sim 10$ MeV. We  regard the high energy group as
the 3-$\hbar \omega$ high-frequency vibrational mode\cite{BM2}. 
In the low-energy group, 
the sharp peak at $E \approx 4.15$ MeV is most prominent, but 
another structure is also seen.
We focus on the low-energy group in the following discussion.

\begin{figure}[htbp]
\begin{center}
\resizebox{0.8\linewidth}{!}{\includegraphics{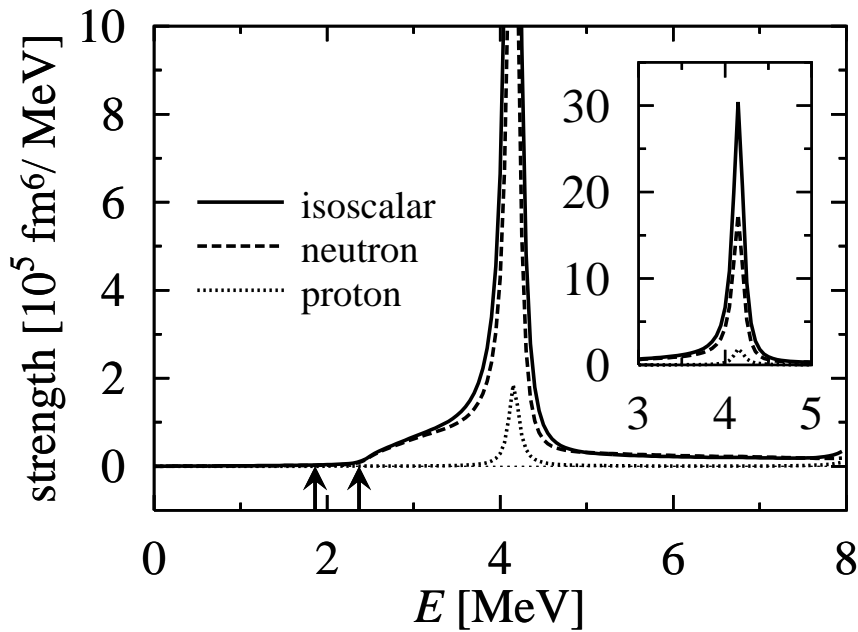}}
\end{center}
\caption{The low-energy part of calculated octupole strengths in $^{84}$Ni.
The isoscalar, neutron and proton strengths are plotted with the solid, dashed and
dotted curves, respectively.  The smoothing parameter
is $\epsilon=0.05$ MeV.  See also the caption of
Fig.\ref{strength.is.ni84}.
}
\label{strength.np.ni84.zoom}
\end{figure}

In Fig.~\ref{strength.np.ni84.zoom}
we show 
magnification of the isoscalar strength function in a low-lying region 
$E=0-8$ MeV, together with the neutron and proton octupole
strength functions. The smoothing parameter 
is chosen  to a smaller value $\epsilon=0.05$ MeV
in this case (and also in most of the following calculations unless
mentioned explicitly).
It is now clear in Fig.~\ref{strength.np.ni84.zoom} that
the low-lying strength consists of two structures.
Apart from the sharp peak at $E=4.16$ MeV, there exists a 
broad bump which emerges above the one-neutron separation energy
(the one- and two-neuron separation energies are $E_{1n,2n}=1.86$ and  2.37
MeV as indicated with arrows).
Although the two structures overlap in the same energy region, 
their characters are clearly distinguished as follows. Firstly, the broad bump 
does not form a well defined peak, and 
we consider it as
a kind of continuum mode where fast neutron decay takes place.
It carries essentially only neutron
strength.  
Secondly, the sharp peak has a small width even though
it is located above the neutron separation energy. It can be regarded 
as a narrow resonance. It is also
distinguished from the neutron mode by the fact that it carries 
sizable proton strength.

\begin{figure}[htbp]
\begin{center}
\resizebox{0.8\linewidth}{!}{\includegraphics{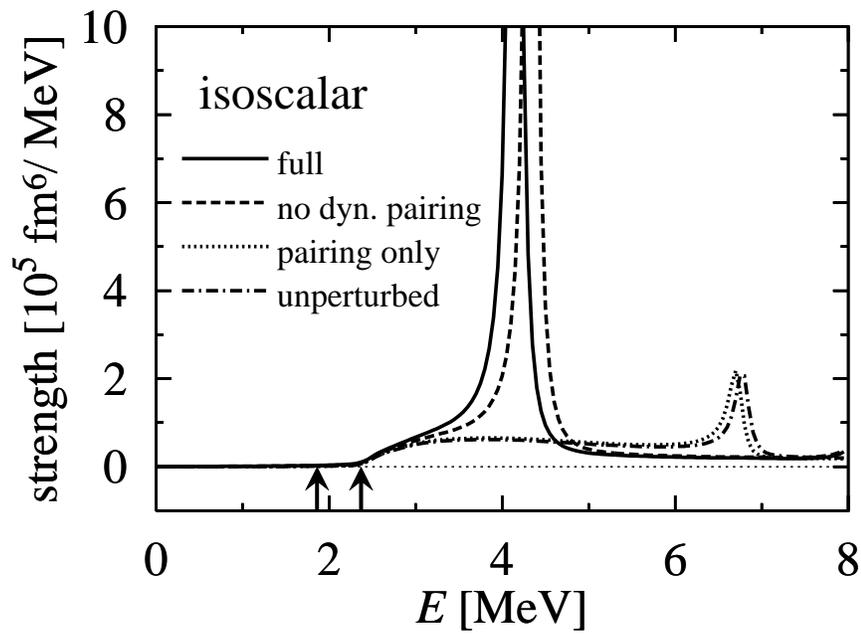}}
\end{center}
\caption{The effects of the residual interactions on the
low-energy part of the isoscalar octupole strength in $^{84}$Ni,
shown in Fig.~\ref{strength.np.ni84.zoom}.
The solid curve
is the full calculation while the dashed and dotted curves are 
the results neglecting the residual pairing and particle-hole
 interactions, respectively. The dot-dashed curve is the unperturbed strength
where both of the
residual interactions are neglected. 
}
\label{strength.is.ni84.resint.zoom}
\end{figure}

In order to give more characterizations, 
we examined how the residual interactions influence these modes.
Namely we performed three calculations where either or both of 
the particle-hole and pairing residual
interactions are switched off. 
Results are shown in Fig.~\ref{strength.is.ni84.resint.zoom}.
The sharp peak disappears
when the particle-hole interaction is neglected, and
it is influenced rather weakly by the pairing residual interaction.
Thus the main origin of the sharp-peak mode is 
 correlation due to the particle-hole residual interaction. 
It suggests that this mode
could be the surface vibrational mode consisting
of low-energy 1-$\hbar\omega$ particle-hole transitions\cite{BM2}.
On the contrary, the broad neutron
mode exists even when the particle-hole residual interaction is
neglected. It has different origin. 

It is useful to evaluate integrated sums of octupole strengths associated with 
the two  modes. 
It is, however, not easy to evaluate them separately
since
the sharp-peak mode and the broad neutron mode overlap in the 
same energy region. Nevertheless we estimate them in the following manner. 
Concerning the broad neutron mode
we integrate the strength functions in an energy region above the
one-neutron separation energy
$E_{1n} < E < E_{1n} + 1.5$ MeV with an interval of 1.5 MeV, 
where the neutron strength dominates and 
the collective vibrational mode barely overlaps.  The boundary energies are
$E_{1n}=1.86$ and $E_{1n} + 1.5=3.36$ MeV in the case of $^{84}$Ni. 
Since the choice of the energy interval 1.5 MeV is rather arbitrary and probably
small to cover the whole strength of the broad neutron mode, 
we expect that there may be an
underestimate by a factor of up to about 2. 
Concerning the strength of the surface vibrational mode, 
we define
an energy interval $E_1 < E < E_2$ where the strengths are integrated by
noticing that this mode carries a proton strength, which is a character 
clearly distinguishable from the broad neutron mode. 
In practice,
we define the boundaries $E_1$ and $E_2$  at which the proton strength
function is 1.0\% of the value at the 
peak energy $E_{peak}$.
In $^{84}$Ni, $E_1$ and $E_2$ are 3.50 and 5.00 MeV, respectively, and there is
no overlap between the two energy intervals. 
In Table \ref{tb:table1}, we list the integrated isoscalar octupole strengths 
of the broad neutron mode and the surface vibrational mode in $^{80-86}$Ni.
The obtained isoscalar octupole strength for the broad neutron mode is
$5.85 \times 10^4 $ fm$^6$ in $^{84}$Ni. Note that the energy-weighted sum of this
strength is 
1.7 percent of the classical isoscalar octupole sum-rule value\cite{BM2,Ring-Schuck}
$S_{cl}^{\rm EW}=
{147 \over 4\pi}{\hbar^2 \over 2m}\left(N\left<r^{4}\right>_n
+Z\left<r^{4}\right>_p \right).$
The isoscalar strength of the surface vibrational mode is $6.74 \times 10^5$ fm$^6$
in the same nucleus, and it exhausts 29 percent of the classical sum rule value.
The strength of the broad neutron mode is smaller than that of the surface 
vibrational mode by about a factor of ten.

\subsection{Transition densities of low-lying octupole modes in $^{84}$Ni}

The natures of the 
sharp-peak mode and the broad neutron mode become more
evident by looking into  transition densities.
We here analyze three kinds of transition densities
\begin{eqnarray}
\rho^{tr}_{iq}(\vecr) &=& \bra{\Phi_i}\sum_\sigma\psi_q^\dag(\vecr \sigma)
\psi_q(\vecr \sigma )\ket{\Phi_0}=
Y_{LM}^*(\hat{\vecr}) \rho^{tr}_{iqL}(r), \\
P^{pp}_{iq}(\vecr) &=& \bra{\Phi_i}\psi_q^\dag(\vecr\uparrow)
\psi_q^\dag(\vecr\downarrow)\ket{\Phi_0}=
Y_{LM}^*(\hat{\vecr}) P^{pp}_{iqL}(r), \\
P^{hh}_{iq}(\vecr) &=& \bra{\Phi_i}\psi_q(\vecr\downarrow)
\psi_q(\vecr\uparrow)\ket{\Phi_0}=
Y_{LM}^*(\hat{\vecr}) P^{hh}_{iqL}(r),
\end{eqnarray}
where $\rho^{tr}_{iq}(\vecr)$ is the usual particle-hole transition density
while $P^{pp}_{iq}(\vecr)$ and $P^{hh}_{iq}(\vecr)$ are the transition densities for
particle-pair and hole-pair, respectively, for either neutrons or protons 
($q=n,p$). They are evaluated as
\begin{eqnarray}
\rho^{tr}_{iqL}(r) &= & -{C \over \pi r^2}{\rm Im}\delta\rho_{qL}(r,\omega_i), \\
P^{pp}_{iqL}(r)&=&{C\over 2\pi r^2}{\rm Im}
(\delta\rhot_{+,qL}(r,\omega_i) -\delta\rhot_{-,qL}(r,\omega_i) ), \\
P^{hh}_{iqL}(r)&=&{C\over 2\pi r^2}{\rm Im}
(\delta\rhot_{+,qL}(r,\omega_i) +\delta\rhot_{-,qL}(r,\omega_i) )
\end{eqnarray}
in terms of the solutions of the linear response equation (\ref{rpa}).  
Here $C$ is a normalization constant, which is fixed so that
the transition amplitude
$M_{iqL}=  \int v^{ext}_{qL}(r)\rho^{tr}_{iqL}(r) r^2 dr$ 
gives the integrated isoscalar octupole strength of a mode
under consideration by the standard definition 
$B({\rm IS3})=7|\sum_q M_{iqL}|^2$.

\begin{figure}[htbp]
\begin{center}
\resizebox{0.6\linewidth}{!}{\includegraphics{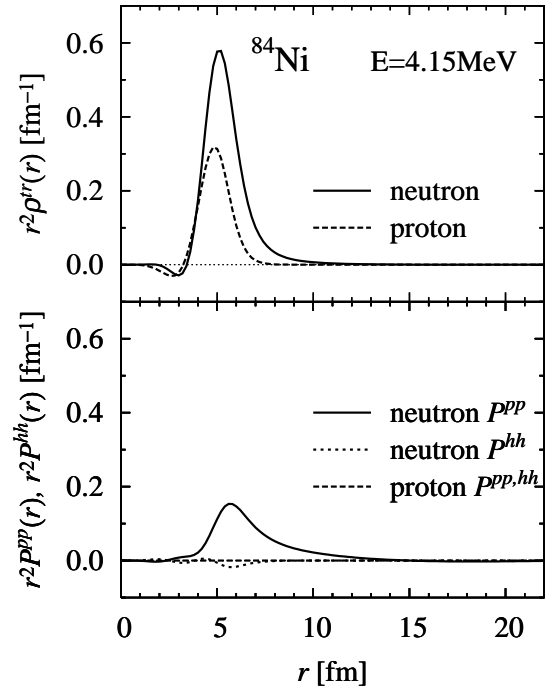}}
\end{center}
\caption{The transition densities evaluated at $E=4.15$ MeV for 
the octupole response of $^{84}$Ni (cf. Fig.~\ref{strength.is.ni84.resint.zoom}).
In the upper panel shown is the particle-hole transition
density $r^2\rho^{tr}_{iqL}(r)$ weighted with the volume element $r^2$.
The neutron and
proton amplitudes are plotted with the solid and dashed curves, respectively.
The particle-pair transition density $r^2P^{pp}_{iqL}(r)$
and the hole-pair transition density $r^2P^{hh}_{iqL}(r)$ of neutrons are shown
in the bottom panel with the solid and dotted curves, respectively.
$P^{pp}_{iqL}(r)$ and $P^{hh}_{iqL}(r)$ for protons (the dashed curves) are exactly zero. 
}
\label{r2trdens.ni84.E4.15}
\end{figure}

In Fig.~\ref{r2trdens.ni84.E4.15},
we show the transition densities associated with the surface
vibrational mode. 
We evaluate them at $E$ = 4.15 MeV, approximately at the peak energy.
It is seen in Fig.~\ref{r2trdens.ni84.E4.15} that the 
particle-hole transition densities 
$\rho^{tr}_{iqL} (r)$ of both neutrons and protons exhibit large and in-phase amplitudes 
at around the nuclear surface (the calculated matter r.m.s. radius of $^{84}$Ni is 4.28 fm).
We thus confirm  that the mode is typical of the surface vibration in which
neutrons and protons give coherent contribution.
It is seen also that the amplitude of the 
particle-hole transition density $\rho^{tr}_{iqL} (r)$ 
is significantly larger than those of the 
particle-pair transition density $P^{pp}_{iqL} (r)$ and the
hole-pair transition density $P^{hh}_{iqL} (r)$
both for neutrons and protons.
($P^{pp}_{iqL} (r)=P^{hh}_{iqL} (r)=0$ for protons is a trivial
consequence of the zero proton pairing gap $\Delta_p=0$.) 
The dominance of the
particle-hole amplitude is consistent with the observation that the
mode is generated by the particle-hole residual interaction (cf. Sec.\ref{sec:strength}).

\begin{figure}[htbp]
\begin{center}
\resizebox{0.6\linewidth}{!}{\includegraphics{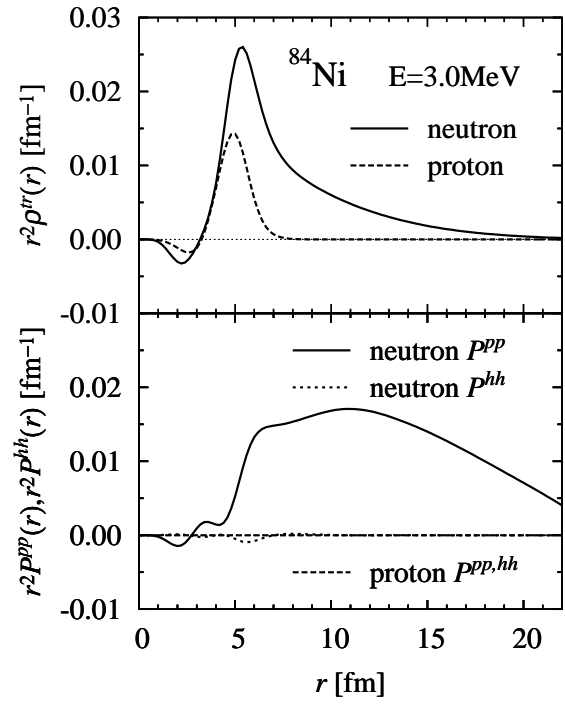}}
\end{center}
\caption{The same as Fig.~\ref{r2trdens.ni84.E4.15} but for the 
transition densities evaluated at $E=3.0$ MeV. 
}
\label{r2trdens.ni84.E3.0}
\end{figure}

Figure \ref{r2trdens.ni84.E3.0} is
the transition densities of the broad neutron mode, evaluated at $E=3.0$ MeV.
They have
characters different from those of the surface vibrational mode.
We observe here two distinct features.
Firstly, 
the particle-pair transition density $P^{pp}_{iqL}(r)$ of neutrons is
significantly larger than 
the neutron particle-hole transition density $\rho^{tr}_{iqL}(r)$ 
in the nuclear exterior region.
The ratio between  the two transition densities 
is approximately a factor of 3 at $r=10$ fm, and a factor of
$\sim 10$ at $r=15$ fm.
This indicates that the mode is characterized, 
especially in the external region,
by motion of neutron pairs rather than by particle-hole excitations.
In other words, the neutron pair correlation is the main character
of this mode.
Secondly, both the particle-hole and particle-pair transition densities 
of neutrons 
exhibit a very long tail extending to $r \sim > 20$ fm, especially
in the particle-pair transition density. This indicates that 
neutrons in weakly bound and continuum orbits
participate in forming this mode. 
In addition to these two features, we observe also that
the proton amplitude of the particle-hole transition
density is considerably smaller than that of neutrons.
This is  in accordance with the dominance of 
the neutron strength already observed in the previous subsection.

\begin{figure}[htbp]
\begin{center}
\resizebox{0.6\linewidth}{!}{\includegraphics{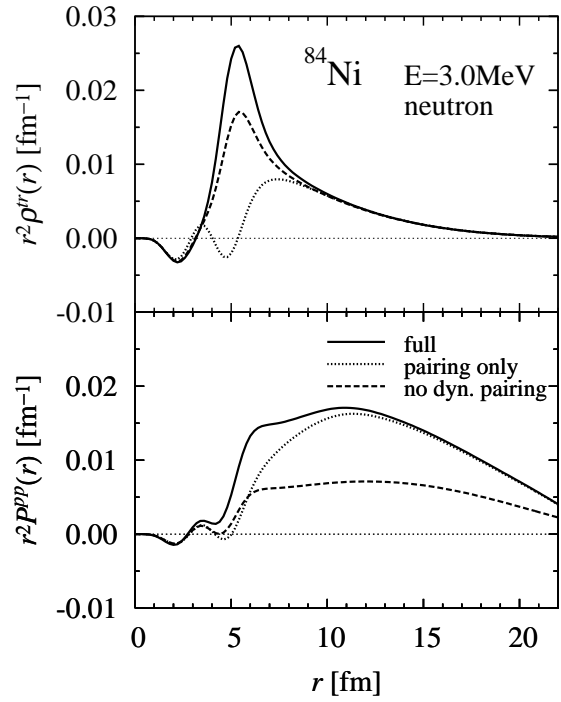}}
\end{center}
\caption{The neutron transition densities $r^2\rho^{tr}_{iqL}(r)$ and $r^2P^{pp}_{iqL}(r)$
of the neutron mode, evaluated 
at $E=3.00$ MeV. Here the effects of the residual interactions are shown.
The dotted curve is the case where the residual particle-hole interaction is neglected 
while the dashed curve is the case where the dynamical pairing effect, i.e. 
the residual pairing interaction, is neglected. The full calculation
is also shown with the solid curve for comparison. See also the caption of 
Fig.~\ref{r2trdens.ni84.E4.15}.
}
\label{r2trdens.ni84.E3.0.pair}
\end{figure}

We  would like to emphasize that the residual pairing interaction 
playing role in the QRPA equation brings the correlation 
 to this mode, and that the pairing mean-field alone is not sufficient.
This is seen in the difference between the solid and dashed lines in
Fig.~\ref{r2trdens.ni84.E3.0.pair},
where the dynamical pairing effect, i.e.,  the 
RPA correlation due to the residual pairing interaction, is either
included or neglected. It is seen that
the dynamical pairing effect enhances the particle-pair transition density 
by a factor of $\sim$ 2.
This indicates significant configuration mixing effect originating
from the residual pairing interaction.
The important role of the pair correlation is seen
even when the particle-hole residual interaction is 
neglected (the dotted curve in Fig.~\ref{r2trdens.ni84.E3.0.pair}).

\begin{figure}[htbp]
\begin{center}
\resizebox{0.6\linewidth}{!}{\includegraphics{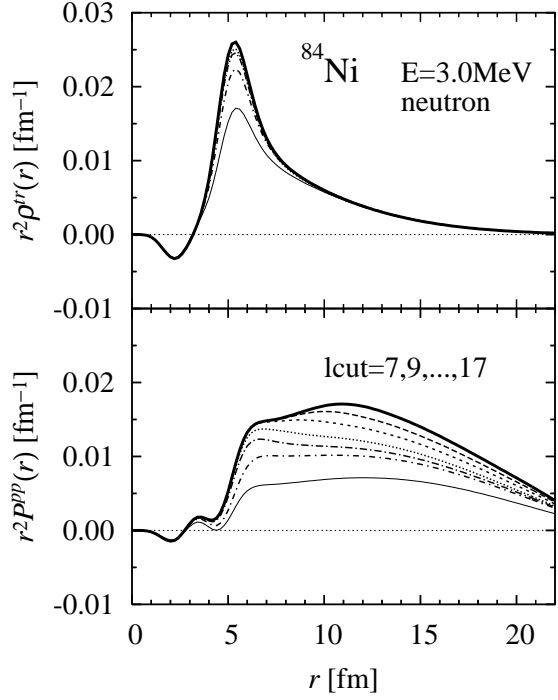}}
\end{center}
\caption{The dependence of the neutron transition densities of the neutron mode, 
evaluated at $E=3.0$ MeV, on
different values of $l_{cut}=7,9,\cdots,17 \hbar$. The upper and lower panels
are for the particle-hole and particle-pair transition densities, $r^2\rho^{tr}_{iqL}(r)$ 
and $r^2P^{pp}_{iqL}(r)$, respectively.
The transition densities obtained neglecting the dynamical 
pairing effect are also plotted with the thin solid curve for comparison. 
}
\label{r2trdens.ni84.E3.0.lcut}
\end{figure}

The large dynamical pairing effect is analyzed in more details.
In Fig.~\ref{r2trdens.ni84.E3.0.lcut} we show how the 
transition densities of the neutron mode changes if we include
only a part of neutron single-particle orbits 
with lower orbital angular momenta $l$, i.e., including only orbits up to 
a smaller cut-off $l_{cut}$ on $l$. This is the same analysis that we performed
for the soft dipole mode\cite{MMS05}.
It is seen that the convergence of the
particle-pair transition density $P^{pp}_{iqL}(r)$ of neutrons with respect to $l$ is
slow, and orbits with large $l$ contribute coherently to produce the
dynamical pairing effect.
There is sizable contribution  even around 
$l \sim l_{max}=17$.  
Note that the precise treatment of the continuum states, guaranteed in the
continuum QRPA approach, is essential to describe the correlation since
high-$l$ orbits with $l > 4$ are all continuum states.

Based on the above two features, i.e., 
the large effect of the dynamical pairing and the high-$l$ contribution, we can
argue that a neutron pair moving 
in the neutron mode exhibits a spatial correlation
at small relative distance between the paired neutrons. The configuration mixing involving 
high-$l$ orbits to a certain value, say $l_{corr}$,  means a two-particle correlation with
a small opening angle $\theta_{corr} \sim 1/l_{corr}$, and hence 
a large $l_{corr}$ means a spatial correlation
at small relative distance. These are the same features seen in the case of the soft dipole
excitation\cite{MMS05}, and we conclude that the di-neutron correlation appears also
in the neutron mode in the octupole response.

\begin{center}
\begin{table}
\begin{tabular}{c c c c c}
\vspace{0.25cm}
\\
\hline
 & 
$^{80}{\rm Ni}$ &
$^{82}{\rm Ni}$ &
$^{84}{\rm Ni}$ &
$^{86}{\rm Ni}$ \\
\hline
neutron mode  \hspace{33mm}\hfill&&&&\\
$B({\rm IS3})$ \ \   $[10^4  {\rm fm}^6]$ &
1.61 &4.23 & 5.85 & 16.11 \\
$S^{\rm EW}({\rm IS3})/S^{\rm EW}_{cl}$ $[\%]$ &0.7&1.6&1.7& 3.4\\
surface vibrational mode \hspace{16mm}\hfill&&&&\\
$E_{peak}$ &5.56&4.78&4.16&3.84\\
$B({\rm IS3})$ $[10^5  {\rm fm}^6]$ &3.57&4.97& 6.74 &10.02\\
$B({\rm E3})$ $[e^210^4  {\rm fm}^6]$&2.77&3.27&3.87&4.53\\
$S^{\rm EW}({\rm IS3})/S^{\rm EW}_{cl}$ $[\%]$ &25.0&26.9&28.6&33.0\\
$M_n/M_p$ &2.63&2.95&3.27&3.90\\
sum rule and separation energies \hspace{5mm}\hfill&&&&\\
$S^{\rm EW}_{cl}$ $[10^6  {\rm fm}^6 {\rm MeV}]$ &7.92 &8.78 &9.77 & 11.36\\
$E_{1n}$ $[{\rm MeV}]$& 2.79 & 2.54 & 1.86 & 1.32 \\
$E_{2n}$ $[{\rm MeV}]$& 3.84 & 3.26 & 2.37 & 1.37 \\
\hline

\vspace{0.25cm}
\\
\end{tabular}
\caption{The isoscalar and electric octupole strengths $B({\rm IS3})$ and $B({\rm E3})$ of the neutron
mode and the surface vibrational mode. The energy weighted sum $S^{\rm EW}({\rm IS3})$ of
the isoscalar octupole strength in unit of the classical sum-rule value $S^{\rm EW}_{cl}$.
is also listed. The neutron vs proton ratio $M_n/M_p$ of the transition amplitudes
as well as the peak energy is also shown for the surface vibrational mode.
The values of  $S^{\rm EW}_{cl}$ as well as the one- and two-neutron separation
energies $S_{1n}$ and $S_{2n}$ are also tabulated.
\label{tb:table1}}
\end{table}
\end{center}

\subsection{Isotopic dependence of the neutron mode}

\begin{figure}[htbp]
\begin{center}
\resizebox{\linewidth}{!}{\includegraphics{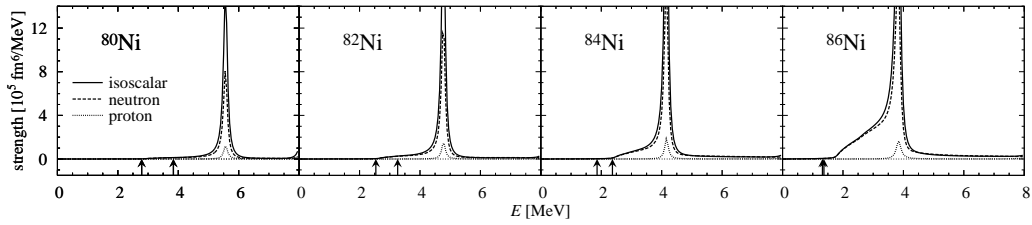}}
\end{center}
\caption{The octupole strength functions obtained for $^{80-86}$Ni. See also the caption of
Fig.~\ref{strength.np.ni84.zoom}
}
\label{strength.np.ni80-86.zoom}
\end{figure}

The strength of the neutron mode increases significantly
as the system approaches toward the neutron drip line.
Figure \ref{strength.np.ni80-86.zoom} shows 
 the strength functions calculated for $^{80,82,84,86}$Ni. The
calculation is the same as that in \S \ref{sec:strength}.
The calculated one-neutron separation energy in these isotopes are
 $E_{1n}$ = 2.79, 2.54, 1.86, and 1.32 MeV in the corresponding order,
 and shown in Fig.\ref{strength.np.ni80-86.zoom} with an arrow.
It is  seen in all the nuclei that there exists above $E_{1n}$  
a broad 
distribution of predominant neutron strength, which corresponds to
the  neutron mode. Clearly the magnitude of the strength 
 increases monotonically with increasing $N$. This is also seen
in the integrated isoscalar strength $B({\rm IS3})$ of this mode, listed in   
 Table \ref{tb:table1}.

\begin{figure}[htbp]
\begin{center}
\resizebox{\linewidth}{!}{\includegraphics{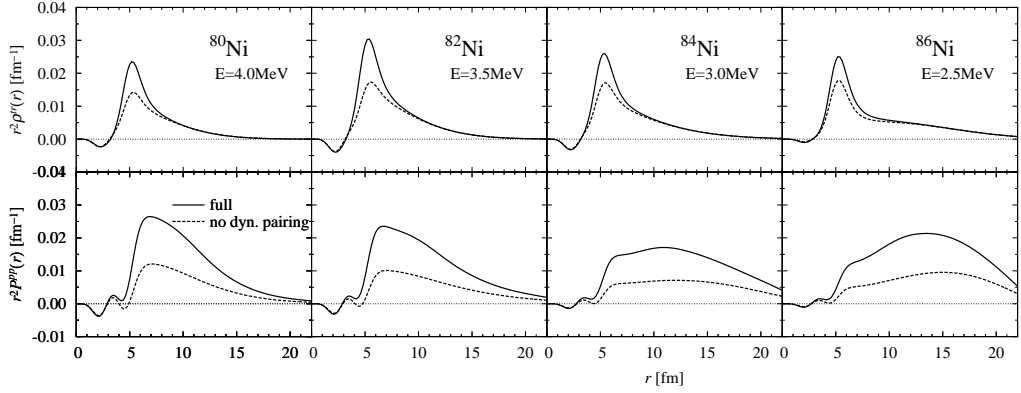}}
\end{center}
\caption{The 
neutron transition densities of the neutron mode obtained for
$^{80-86}$Ni. 
The upper and lower panels
are for the particle-hole and particle-pair transition densities, $r^2\rho^{tr}_{iqL}(r)$ 
and $r^2P^{pp}_{iqL}(r)$, respectively. We calculate the transition densities in each nucleus
at the excitation energy listed in the Figure.
The transition densities obtained neglecting the dynamical 
pairing effect are also plotted with the dashed curves for comparison. 
}
\label{r2trdens.ni80-86}
\end{figure}

The transition
densities of the neutron mode, evaluated at 
$E$ = 4.0, 3.5, 3.0, and 2.5 MeV for $^{80,82,84,86}$Ni, respectively,
are shown in Fig.~\ref{r2trdens.ni80-86}.
In the top panels 
we see that the tail of the particle-hole transition density grows 
as the neutron number increases. 
It is known that such a long tail can be realized for  particle-hole 
transitions from weakly bound orbits, whose wave functions 
have long tail, to continuum orbits with small kinetic energies.
The long tail enhances the octupole strength as
the octupole operator has 
the radial form factor $\propto r^3$ giving a heavy weight at larger
distances. We thus see that
the increase of strength with increasing $N$ originates from
the effect of weak binding of neutrons.

Looking at 
the particle-pair transition densities shown in 
the bottom panels in Fig.~\ref{r2trdens.ni80-86},
we observe that the particle-pair transition
density varies with $N$ more drastically  than the particle-hole 
transition density.
In $^{86}$Ni and also in $^{84}$Ni
 this transition density  does not show exponential decay in the outside
of the nucleus $r \ge 10$ fm, but it rather shows
an oscillatory behaviour with its maximum amplitude 
at around $r\sim 15$ fm far outside the nucleus.
 This suggests significant {\it emission of
a  neutron-pair}  from the nucleus. This is of course related to the
feature that the two-neutron separation 
energy $E_{2n}$ decreases from $E_{2n} = 3.84$ MeV in 
$^{80}$Ni to a small value 1.37 MeV in $^{86}$Ni. 
The di-neutron correlation in the neutron mode 
becomes more significant as we approach the neutron drip-line.

\subsection{Model dependence}

Let us now investigate how our predictions on the
neutron mode depends on the model parameters. For this purpose,
we shall compare the above results with those obtained with another Skyrme parameter set
SkM$^*$\cite{SKMstar}, and those with a model adopting a Woods-Saxon potential \cite{MMS05}
instead of the
Skyrme-HFB self-consistent mean-fields. 
When calculating the octupole response for these models, 
we use the same mixed-type DDDI, but the interaction parameter $V_0$ is adjusted separately
to reproduce the average pairing gap of neutrons in stable nuclei as in the SLy4 case: 
$V_0=-265$ and $-280$MeV fm$^{-3}$
for SkM$^*$ and WS, respectively. 
The obtained average neutron pairing gap in $^{84}$Ni is 1.349, 0.797 and 0.569 MeV
for the Woods-Saxon, SLy4 and SkM$^*$ models, respectively.

\begin{figure}[htbp]
\begin{center}
\resizebox{0.8\linewidth}{!}{\includegraphics{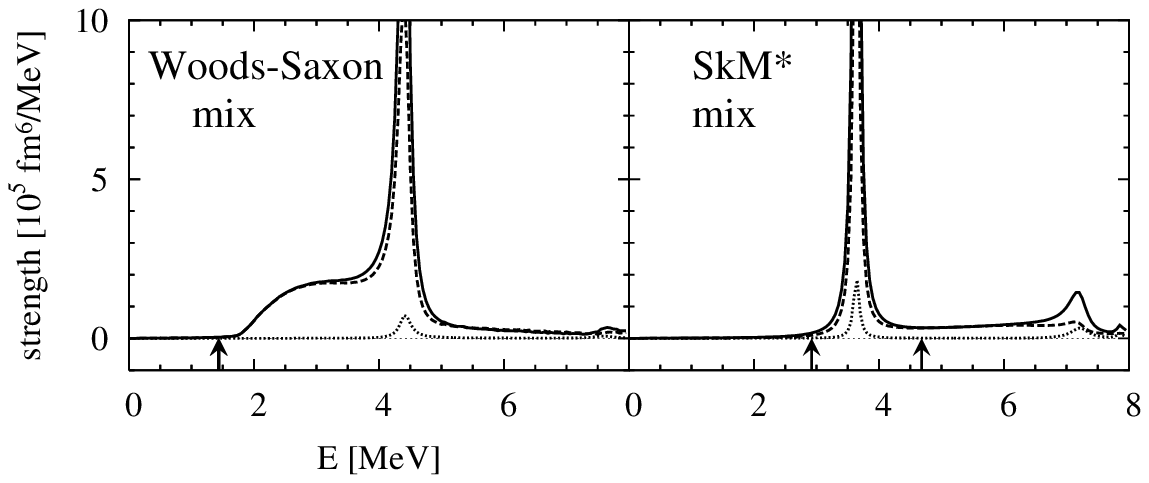}}
\resizebox{0.8\linewidth}{!}{\includegraphics{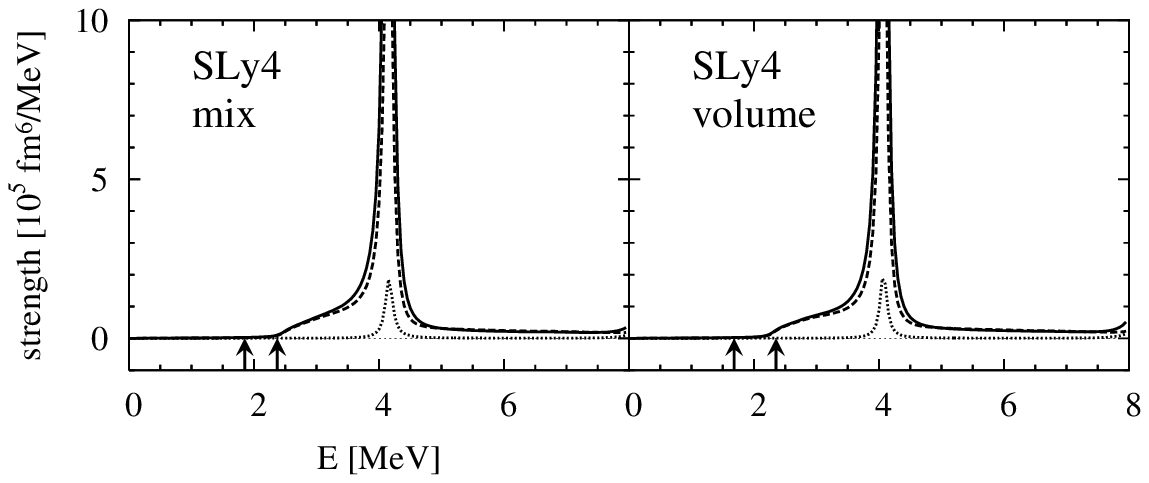}}
\end{center}
\caption{
The octupole strength functions in $^{84}$Ni, obtained with the 
Woods-Saxon model (left top) and the Skyrme HFB model with 
SkM$^*$ (right top). The mixed-type DDDI is used in both cases. 
The bottom panels show the strength functions 
for the Skyrme parameter set SLy4 but with different
pairing interaction: the mixed-type DDDI (bottom left) and the volume-type 
density-independent delta interaction (bottom right).
See the caption of Fig.~\ref{strength.np.ni84.zoom} for
the definition of curves.
}
\label{strength.SkMWS}
\end{figure}

\begin{figure}[htbp]
\begin{center}
\resizebox{0.6\linewidth}{!}{\includegraphics{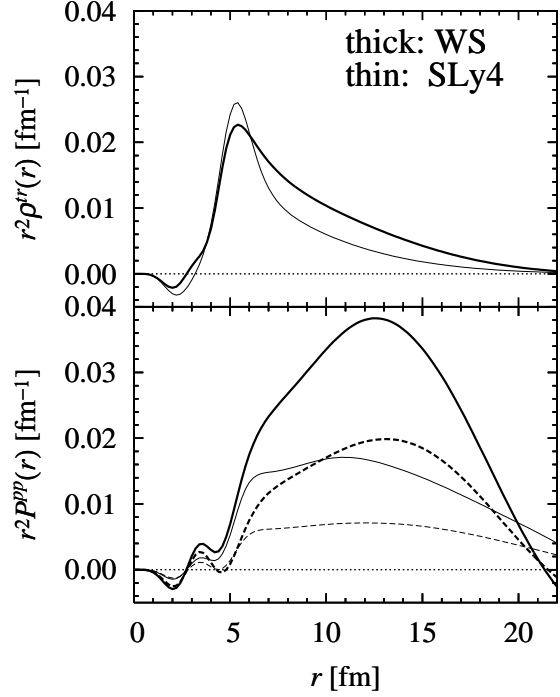}}
\end{center}
\caption{
The comparison of the neutron transition densities 
$r^2\rho^{tr}_{iqL}(r)$ 
and $r^2P^{pp}_{iqL}(r)$ of the neutron mode in $^{84}$Ni
obtained with the Woods-Saxon model (the thick curves)  vs. those with the 
Skyrme HFB with SLy4 (the thin curves). The mixed-type DDDI is used and
the excitation energy is $E=3.0$ MeV in both cases. 
The dashed curves in the bottom panel represent the results
obtained neglecting the dynamical pairing effect.
}
\label{r2trdens.WS}
\end{figure}

The octupole strength function in $^{84}$Ni obtained for the SkM$^*$ and Woods-Saxon
models are shown and compared with that for SLy4 in Fig.\ref{strength.SkMWS}.
It is seen that the neutron mode depends rather sensitively
on the models. 
The Woods-Saxon model produces a significantly larger
strength than SLy4 while it
is smallest for SkM$^*$ among the three models.
Figure \ref{r2trdens.WS} is a comparison of the 
transition densities of the soft neutron mode of the
Woods-Saxon model and that of the SLy4, both evaluated at $E=3.0$ MeV.
(The transition densities for SkM$^*$ is not shown here
as the strength of the neutron mode itself is weak.)
The Woods-Saxon model exhibits enhanced amplitudes of the transition 
densities,
especially of the particle-pair transition density, compared
with the SLy4 model.

\begin{figure}[htbp]
\begin{center}
\resizebox{0.6\linewidth}{!}{\includegraphics{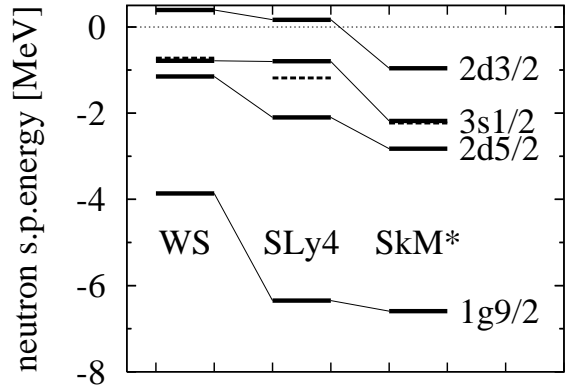}}
\end{center}
\caption{
The neutron single-particle energies for $^{84}$Ni in the 
Woods-Saxon model and the Skyrme Hartree-Fock model with 
SLy4 and SkM$^*$. The dashed line indicates the neutron
Fermi energy.
}
\label{spe}
\end{figure}

We can relate the model dependence  to differences in 
the single-particle levels and the Fermi energy of neutrons. 
The neutron Fermi energy in $^{84}$Ni is $\lambda_n=-0.722, -1.183$ and 
$-2.338$ MeV
for the Woods-Saxon, SLy4 and SkM$^*$ models, respectively.
The neutron single-particle levels are shown in Fig.\ref{spe}.
Clearly the last neutrons are less (more) bound in the case of
the Woods-Saxon model (the SkM$^*$ model). Thus the 
 model dependence of the low-lying octupole strength can be
explained in terms of
the weak-binding effect which increases the strength of the neutron mode.

\begin{figure}[htbp]
\begin{center}
\resizebox{0.6\linewidth}{!}{\includegraphics{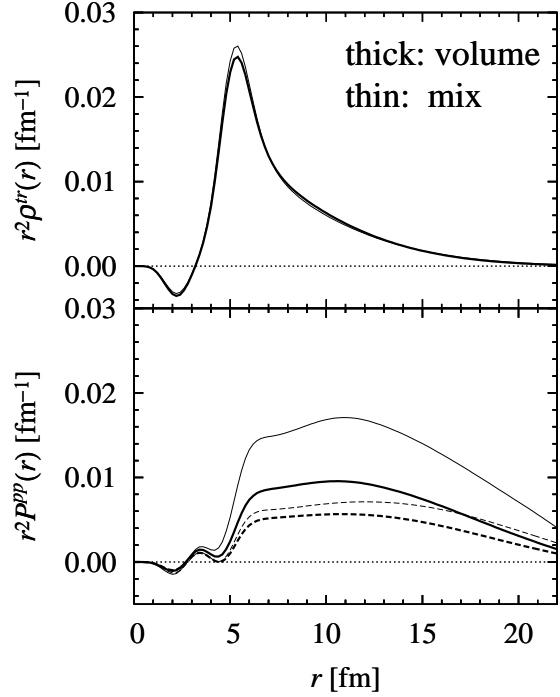}}
\end{center}
\caption{
The dependence of the neutron transition densities 
$r^2\rho^{tr}_{iqL}(r)$ 
and $r^2P^{pp}_{iqL}(r)$ of the neutron mode on the pairing interaction.
We compare results obtained with the mixed-type DDDI (the thin curves) 
and the volume-type density-independent
delta interaction (the thick curves). The excitation energy is $E=3.0$ MeV in both cases.
The Skyrme parameter set SLy4 is used. 
}
\label{r2trdens.mixvol}
\end{figure}

We can speculate that the
neutron mode depends also on the effective pair interaction.
Let us examine explicitly
 dependence on the effective pair interaction.
For this purpose we compare results obtained with the mixed 
type DDDI with another calculation using 
the density-independent delta interaction,
which is defined by Eq.(\ref{DDDI}) but with $\eta=0$.
(The same 
SLy4 is used as the Skyrme parameter set.)
The strength 
$V_0=-215$MeV fm$^{-3}$ is adjusted in the same way as for the
mixed type DDDI, i.e. to reproduce the average neutron paring
gap in stable Ni isotopes.
The density independent delta interaction is often
called the volume pairing. The volume pairing
produces average neutron pairing gap of 0.467 MeV in
$^{84}$Ni, slightly
smaller than that in the mix-type DDDI. 
As far as the octupole strength function is concerned, there
is no significant dependence on the type of the pairing interaction
as we see from the comparison in
Fig.\ref{strength.SkMWS}. There is, however, large
difference in the 
particle-pair transition density $P^{pp}_{iqL}(r)$ of neutrons, as
shown in Fig.~\ref{r2trdens.mixvol}. The amplitude
of the particle-pair transition density $P^{pp}_{iqL}(r)$ 
in the volume pairing case is smaller by a factor of $\sim 2$ than
that of the mixed-type DDDI. It is seen also that the dynamical paring effect
is significantly smaller in the volume pairing case. 
These results indicate that the di-neutron correlation in the 
neutron mode is  sensitive to the choice of 
the effective paring interaction:
effective pairing interactions such as the mixed-type DDDI that has
stronger interaction strength at low densities give stronger di-neutron
correlation in the neutron mode. 
This feature is in agreement with our previous finding for the soft
dipole excitation\cite{MMS05}.

\subsection{Comparison with soft dipole excitation}

The analyses in the preceding subsections 
revealed that the
weak binding effect and the di-neutron correlation in the
octupole neutron mode are similar 
to those seen in the soft dipole excitations\cite{MMS05}.
In this subsection let us make 
a more explicit and quantitative comparison between the dipole
and octupole cases. For this purpose we calculate the electric dipole response 
using the same Skyrme-HFB + continuum QRPA model.
(The calculations 
in our previous work\cite{MMS05} is based on the Woods-Saxon model, and
not suitable for direct comparison with the present calculations.)

\begin{figure}[htbp]
\begin{center}
\resizebox{\linewidth}{!}{\includegraphics{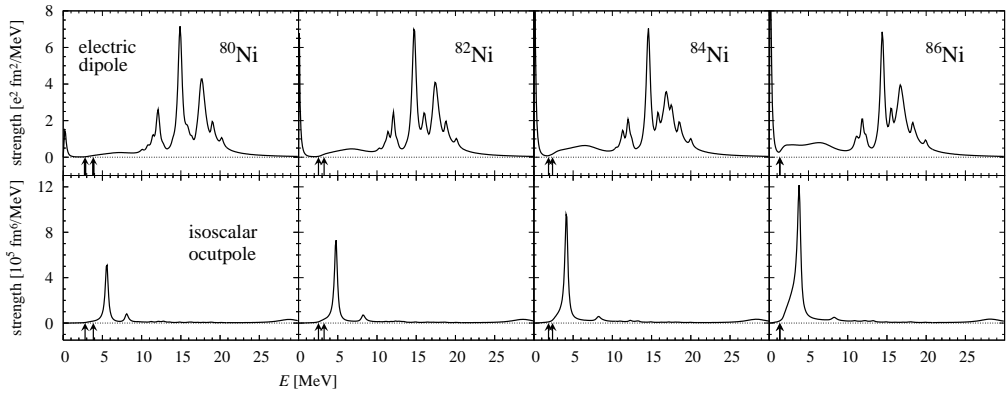}}
\end{center}
\caption{The electric dipole strength function $dB({\rm E1})/dE$ in $^{80-86}$Ni, 
shown in the top panels, obtained with the
Skyrme parameter set SLy4 and the mixed-type DDDI, 
compared with the isoscalar octupole strength function $dB({\rm IS3})/dE$ shown in the 
bottom panels.
The smoothing parameter is $\epsilon=0.2$ MeV.
}
\label{strength.L13.ni80-86}
\end{figure}

The calculated electric dipole strength function $dB({\rm E1})/dE$ is 
shown in Fig.\ref{strength.L13.ni80-86}.
In the same energy region as the octupole neutron mode
(i.e. just above the one-neutron separation energy), 
there emerges a broad distribution of the E1 strength corresponding to
the soft dipole excitation. It is seen also that
the monotonic increase of the strength with $N$
is in parallel with that of the octupole
neutron mode. 
If we evaluate the $B({\rm E1})$ strength integrating within an
energy interval $[E_{1n}, E_{1n}+5{\rm MeV}]$   above the one-neutron
separation energy (as done in Ref.\cite{MMS05}), the obtained E1 strength is
$B({\rm E1}; 0_{gs}\rightarrow 1^{-})=0.83, 1.59, 2.23, 3.33$ 
$e^2{\rm fm}^2$ for
$^{80,82,84,86}$Ni, respectively. 
The energy weighted sum in the same energy interval is 1.9, 3.3, 4.0 4.8\% 
of the classical (TRK) sum rule value for $^{80,82,84,86}$Ni.
The E1 strength and the energy weighted sum in
a smaller energy interval $[E_{1n}, E_{1n}+1.5 {\rm MeV}]$ are 
0.11, 0.24, 0.35, 0.87 $e^2{\rm fm}^2$
and 0.15, 0.30, 0.36 0.68\% for the same isotopes. 
If we compare the strengths of the
octupole  neutron mode and the soft dipole mode in terms of
the fraction of the energy weighted strength to the classical
sum rule,  1.7\% for the octupole neutron mode and 4.0\% for the soft dipole
in $^{84}$Ni (0.4\% if the same energy interval $[E_{1n}, E_{1n}+1.5{\rm MeV}]$ is used)
are comparable. 

\begin{figure}[htbp]
\begin{center}
\resizebox{0.6\linewidth}{!}{\includegraphics{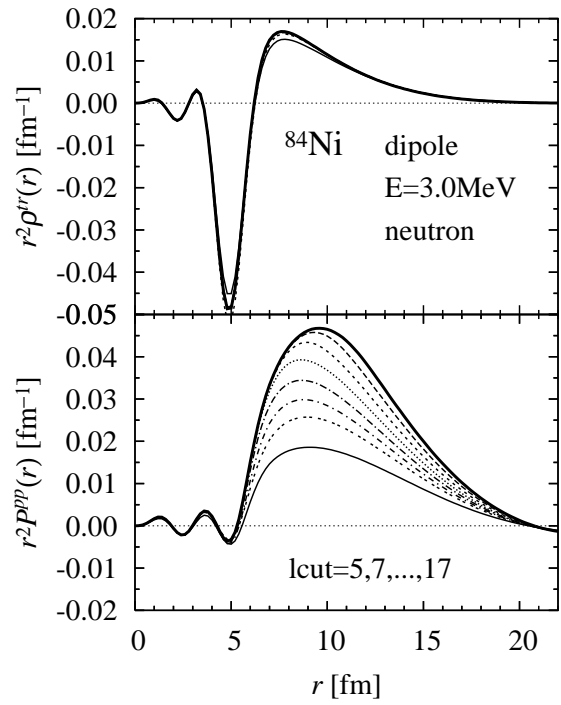}}
\end{center}
\caption{The same as Fig.\ref{r2trdens.ni84.E3.0.lcut} but for the soft dipole excitation. 
Here the angular momentum cut-off is $l_{cut}=5,7,9,\cdots 17\hbar$. The
transition densities are evaluated at $E=3.0$ MeV using the mixed-type DDDI and SLy4. 
}
\label{r2trdens.L1.ni84.lcut}
\end{figure}

In Fig.~\ref{r2trdens.L1.ni84.lcut}, we show the transition
densities of the soft dipole mode. The transition densities
are evaluated  at the same excitation energy $E=3.0$ MeV
and in the same energy interval 
[$E_{1n}, E_{1n}+1.5$ MeV] as is done for the octupole neutron mode.
Comparing the transition densities of the octupole (Fig.\ref{r2trdens.ni84.E3.0.lcut}) and
dipole (Fig.\ref{r2trdens.L1.ni84.lcut}) modes, both are similar in that
the particle-pair transition density $P^{pp}_{iqL}(r)$ is
most dominant, and also 
that the dynamical pairing effect and coherent contributions
of the high-$l$ orbits play significant role to enhance
the particle-pair transition density. 
If we compare the absolute magnitudes of the transition densities,
it is seen that the magnitude of
the particle-pair and particle-hole transition densities
of the octupole neutron mode is smaller than those of
the soft dipole mode by a factor of $\sim$ 2
at around $r=10$ fm where the particle-pair amplitude is
largest. Apart from this difference, 
the significance of the di-neutron correlation is
comparable in both cases.

The most noticeable difference between the octupole and dipole response is
that in the octupole response the neutron mode  overlaps
with the surface vibrational mode present in the same energy region,
and the strength is overwhelmed by the latter. In contrast, the
soft dipole excitation in the dipole response is well separated from
the other mode of excitation, the giant dipole resonance. This may 
make it more difficult to identify experimentally the octupole  neutron
mode.

\subsection{Surface vibrational mode}

Let us mention briefly the octupole surface vibrational mode,
corresponding to the sharp peaks around $E\approx 5.6-3.8$ MeV in
$^{80-86}$Ni.
The peak energies as well as the isoscalar and electric 
octupole strengths associated with the surface vibrational mode 
are listed in Table~\ref{tb:table1}. 
The most noticeable feature is that  the isoscalar 
strength increases
 steeply with increasing $N$ from 80 to 86
approximately  by a factor of three. We also see that
the neutron vs. proton ratio $M_n/M_p$
of the transition amplitudes 
increases more steeply than the 
the nominal ratio $N/Z$. The double ratio
$(M_n/M_p)/(N/Z)$ increases from 1.41 to 1.88 with increasing $N$.
These indicate that
the enhanced collectivity of the surface vibrational
mode is due mainly to neutron contributions.  
Note that the ratio of the energy weighted
sum of the isoscalar strength to
the classical sum rule value $S_{cl}^{EW}$ 
stays constant around 25-33 \%. This is partly 
because the increase of
the strength is compensated by the decrease of the excitation energy
(see Table~\ref{tb:table1}), and 
partly because the sum rule value  $S_{cl}^{EW}$ 
itself increases with $N$ due to increasing
radial expectation value $\left<r^4\right>_n$
of neutrons. This suggests that
the increase of the isoscalar and neutron
strength of the surface vibrational mode is regarded as a 
kind of softening
caused by the weak binding of neutrons.

The enhanced collectivity of the octupole vibrational
mode in neutron-rich nuclei close to the drip-line is
pointed out in
Ref.\cite{HSZ01},  which however analyzed 
a doubly-closed-shell nucleus $^{60}$Ca 
using the Skyrme-HF plus continuum
RPA without the pair correlation.  Our results
suggest that the enhanced collectivity is generally seen
in nuclei near the neutron drip line.
A large deviation of the $M_n/M_p$ ratio from the nominal 
ratio $N/Z$ is pointed out 
 in a Skyrme-BCS+ QRPA calculation\cite{KBG00}
for the neutron-rich oxygen isotope $^{24}$O, 
but with much smaller deviation in less neutron-rich isotopes
$^{18-22}$O.  
We refer also to Ref.\cite{Yam05}, which 
discuss enhanced collectivity due to the weak binding
effects in the case of
the low-lying quadrupole vibrational mode.

\section{Conclusions}

We have investigated the low-lying 
octupole excitations of the neutron-rich Ni isotopes beyond the $N=50$ shell closure
using the continuum QRPA based on the Skyrme Hartree-Fock-Bogoliubov mean-fields.
In addition to the surface vibrational mode of the 1-$\hbar\omega$ character,
a broad strength distribution of predominantly neutron component appears 
just above the neutron separation energy. This broad neutron mode exhibits
the following distinctive features.
(i) The transition amplitude for neutron pair density is significantly
larger than that of the usual particle-hole transition density of neutrons. 
The neutron pair correlation is therefore the most essential aspect characterizing
the neutron mode. 
(ii) The large neutron-pair transition density originates from a coherent contribution 
of the neutron high-$l$  orbits, which points to the presence of 
the spatial correlation among neutrons involved in this mode. 
(iii) The transition densities of the neutron mode display a long tail extending in the
outside of the nucleus. The strength of this mode increases monotonically with
increasing $N$ from 52 ($^{80}$Ni) to 54, 56 $\cdots$. Both indicate
that the mode originates from the weak binding of neutrons.
From these features we conclude that the  
spatial di-neutron correlation  shows up  also in the neutron mode in
 nuclei near the neutron drip-line, similarly to the soft dipole 
mode. 
This supports our expectation 
that soft modes
having the di-neutron character may emerge generically in 
medium mass neutron-rich nuclei near the drip-line. 
It is interesting to investigate this generality
in more details, e.g. by looking into modes with other multipolarities. Such an
analysis is in progress, and will be reported elsewhere.

\section{Acknowledgments}

This work was supported by the Grant-in-Aid for Scientific
Research (Nos.17540244, 20540259) from the Japan Society for the Promotion of
Science, and also by the JSPS Core-to-Core Program, International
Research Network for Exotic Femto Systems(EFES). 
The numerical calculations were carried out on SX5 at Research Center
for Nuclear Physics in Osaka University, and SX8 at Yukawa Institute for
Theoretical Physics in Kyoto University.


\clearpage


\end{document}